\documentclass[sigconf]{acmart}
\setcopyright{none}
\renewcommand\footnotetextcopyrightpermission[1]{}

\usepackage{booktabs}

\begin{document}

\title{Deforking the World of Code:\\
       A Project-Provenance Map that Recovers Cross-Forge Fork
       Families that Platform Graphs Cannot See}

\author{Audris Mockus}
\affiliation{%
  \institution{University of Tennessee, Knoxville}
  \city{Knoxville}\state{TN}\country{USA}}
\email{audris@utk.edu}

\begin{abstract}
Forks share git history, so a commit surfaces in many repositories and any
spread- or popularity-based measure over raw repositories is inflated by orders
of magnitude. We release a curated \emph{deforking} map for the World of Code
(WoC) version \texttt{V2604}: \texttt{p2PFull}, which collapses every raw
repository $p$ into the deforked project $P$ to which it belongs, built from the
global shared-commit relation ($51.79$M shared-commit groups) via a hub-node
star encoding and parallel Louvain clustering, plus capped variants
(\texttt{cap250}/\texttt{cap500}) that bound mega-cluster size. The naive
shared-history union over-merges: the project graph welds unrelated software
into giant clusters (largest uncapped cluster $443{,}010$ repositories, bridged
by shared-commit groups as large as $267{,}200$), for the same structural reason
author-identity graphs do. A cheap size cap removes the boilerplate-hub bridges;
a structural-bridge diagnostic, the cut that dissolved the analogous author
mega-cluster, run here but \emph{deliberately not applied}, shows the post-cap
residual is \emph{genuine vendored history}, robust to the cut, so we leave it
intact. We validate the map against GitHub's declared fork graph reconstructed
from GHArchive \texttt{ForkEvent}s, finding $99.01\%$ edge agreement conditional
on both repositories being in WoC. Disagreements fall into two classes: a
completeness byproduct (edges GitHub asserts but WoC has not ingested) and the
central contribution, \emph{WoC-only fork families} that GitHub's platform graph
cannot represent, including $5.53\%$ multi-forge families and $1.54\%$ whose fork
root is not on GitHub. We additionally release a refreshed fork-exclusion list
($134.1$M children, $3.4\times$ the GHTorrent-era $39.5$M) and a detached-fork
inventory ($455{,}550$ hard-detached edges; $240{,}441$ genuine independent
origins). All artifacts are a self-contained, independently hosted replication
package keyed to the WoC \texttt{V2604} collection.
\end{abstract}


\maketitle

\section{Introduction}
\label{sec:intro}
Mining software repositories at global scale must contend with forking: a single
upstream history is copied into thousands of derived repositories, so a commit,
a file, or an author appears to span far more ``projects'' than it truly does.
Left uncorrected, every spread-, reach-, and popularity-based signal (author
breadth, project size, dependency fan-out) is inflated by the fork multiplicity.
\emph{Deforking}, the recovery of the canonical project a repository belongs to,
is therefore a prerequisite construction for any global-scale analysis, and it is
exactly as error-prone as author-identity disambiguation: the obvious
shared-history union over-merges catastrophically, and the fix requires the same
precision discipline.

This paper releases a deforking map for the World of Code (WoC)
\texttt{V2604} collection, extending the original shared-commit deforking
approach~\cite{forks20} to the current collection, and documents both its
construction and its external validation. Our contributions are the released artifacts and four findings about
them, recorded as a living experiment log (Exps.~D1--D5):
\begin{itemize}
\item \textbf{The map and its capped variants} (\texttt{p2PFull},
  \texttt{cap250}/\texttt{cap500}): a repository$\rightarrow$project provenance
  map over $51.79$M shared-commit groups, mega-cluster controlled (Exp.~D1).
\item \textbf{A structural diagnostic} showing the post-cap residual is genuine
  shared (vendored) history rather than a clustering artifact, robust to the cut
  (Exp.~D2), with its measured effect on the spread signal (Exp.~D3).
\item \textbf{External validation} against GitHub's declared fork graph from
  GHArchive \texttt{ForkEvent}s: $99.01\%$ edge agreement conditional on both
  endpoints in WoC, plus a completeness byproduct (Exp.~D4).
\item \textbf{Cross-forge and detached-fork analysis}: the WoC-only fork families
  invisible to any single platform's graph, a refreshed fork-exclusion list, and
  a detached-fork inventory (Exp.~D5).
\end{itemize}
The dataset is a drop-in join key for any WoC-scale study and, because it is
built from observed shared commits rather than declared platform metadata, it
captures fork relationships that span forges and survive history detachment,
the cases platform fork graphs miss.

\section{Related Work}
\label{sec:related}
The closest prior work is the original shared-commit deforking map for the World
of Code~\cite{forks20}. It established the idea this paper builds on: because git
commits are Merkle-chained and practically impossible to reproduce independently,
two repositories that share a commit are almost certainly the same project, so
linking repositories by shared commits and extracting communities recovers fork
families. Working over an earlier WoC version (about $1.9$B commits and $116$M
repositories), it found that the naive shared-commit union produces a
$13.9$M-repository mega-cluster, attacked it by manually removing backup and
multi-project repositories and dropping commits that span more than a thousand
projects, substituted GitHub's ultimate parent for declared forks, and ran
Louvain community detection in R/iGraph (in chunks, to skirt iGraph's long-vector
limit), reducing the largest community to about $355$K repositories. It validated
against later-retrieved GitHub fork parents and against a GitHub-fork-based
deduplication dataset~\cite{spinellis2020dedup}.

We build directly on that approach and extend it in five ways, each documented as
an experiment below. (1)~\emph{Scale and method}: we deform the current WoC
\texttt{V2604} collection ($5.87$B commits, roughly three times the original),
clustering a hub-node star encoding of the shared-commit relation with parallel
Louvain rather than the original's chunked R/iGraph run, which could not hold the
graph in memory at once. (2)~\emph{A deterministic size cap}: dropping
shared-commit groups larger than a cap \textsc{C} inside the graph builder removes
the boilerplate hubs that bridge unrelated families, recall-safely and
reproducibly. This generalizes the original's manual bad-project list and its ad
hoc removal of commits spanning more than a thousand repositories into a single
tunable, in-builder parameter (Exp.~D1). (3)~\emph{A structural diagnosis} of the
residual: a sampled-betweenness analysis shows the post-cap residual is genuine
vendored history (Chromium in Qt in Debian), not a clustering artifact, and so
should be kept rather than cut (Exp.~D2). (4)~\emph{Cross-forge framing}: because
the map clusters by shared commits and not by any forge's fork pointer, it
recovers fork families that span forges and families rooted off GitHub entirely,
which we quantify (Exp.~D5). (5)~\emph{Event-stream validation at full scale}:
rather than slowly polling the GitHub fork API, we reconstruct GitHub's declared
fork graph from the GHArchive \texttt{ForkEvent} stream and cross-check the entire
map against it, turning the disagreement into a usable cross-forge and
ingestion-completeness signal (Exp.~D4).

The declared-fork resources we validate against are platform-metadata graphs.
GHTorrent mirrors GitHub's event API and exposes parent--child fork edges and a
fork-exclusion set~\cite{gousios2012ghtorrent,gousios2013ghtorrent}; GitHub's own
REST/GraphQL fork API offers the same relation live; and the deduplication
dataset~\cite{spinellis2020dedup} packages GitHub's ultimate-parent forks as a
cleaning resource. All are \emph{metadata} graphs: an edge exists only if a fork
was created through a forge's UI, on that forge, within the recorded window.
Kalliamvakou et al.~\cite{kalliamvakou2014promises} document the resulting perils:
many declared forks are empty, never-pushed, or otherwise inactive, and the
recorded chain root need not be the true upstream, which our edge-level partition
(Exp.~D4) quantifies directly. Our map is complementary and commit-based:
following World of Code~\cite{ma2019woc,ma2021world}, we recover the canonical
project from \emph{observed shared commits}, so a fork is detected whenever two
repositories share git objects, regardless of which forge hosts them or whether
the fork button was ever pressed. This commit-based, cross-forge consolidation is
in the spirit of reconstructing a single evolutionary history out of many mirrored
repositories, as in Spinellis's Unix-history work~\cite{spinellis2017unix}.

\section{Deforking the Project Graph}
\label{sec:deforking}

\begin{figure}[t]
\centering
\includegraphics[width=0.84\linewidth]{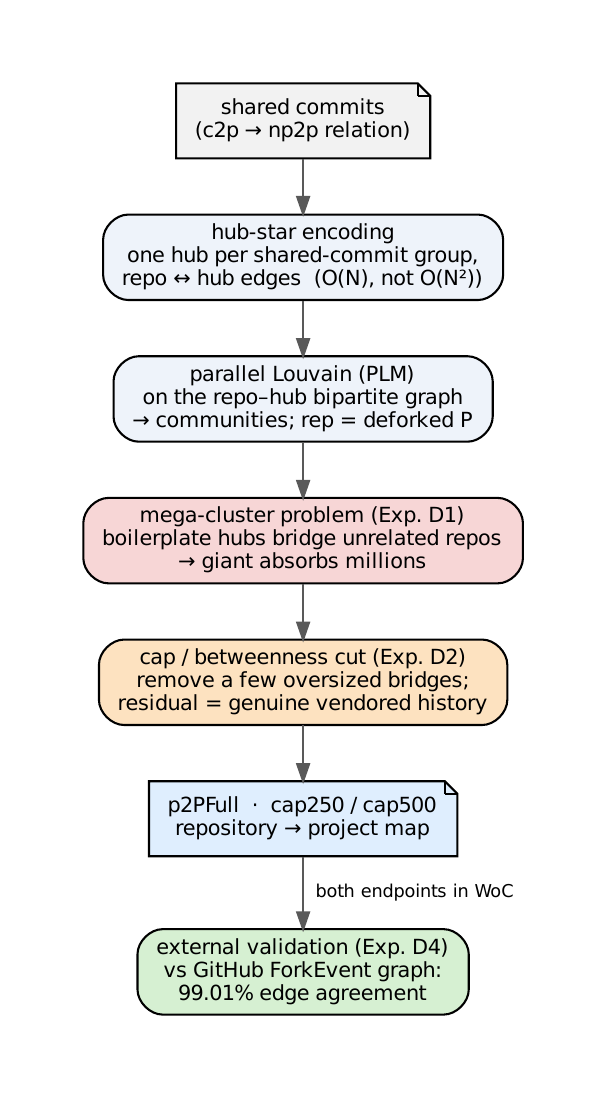}
\caption{Shared-commit deforking. A hub-star encoding makes the shared-commit
relation linear rather than quadratic, and parallel Louvain clusters the
repository--hub graph into deforked projects. Uncapped, boilerplate hubs bridge
unrelated repositories into a mega-cluster that a cap removes, leaving a residual
of genuine vendored history.}
\label{fig:defork}
\end{figure}

Forks share git history, so the same commit appears in many WoC ``projects,'' and
any spread-, reach-, or popularity-based measure read off raw repositories is
inflated by the fork multiplicity: an author of one popular-but-forked repository
would falsely register a spread of thousands of distinct projects, and a file or a
dependency appears to reach far more projects than it truly does. Deforking, the
recovery of the canonical project $P$ (uppercase) a raw repository $p$ belongs to,
is therefore a prerequisite for any spread- or overlap-based signal, and its naive
form over-merges and needs a precision fix (Figure~\ref{fig:defork}).

\paragraph{Flattening commits to a project graph.}
Two repositories belong to the same fork family when they share commits. We
materialize the shared-commit relation as \texttt{np2p}: one row per distinct
set of repositories that share an identical commit, \texttt{count;$p_0$;\dots;$p_N$}
($51.79$M groups in V2604). Rather than connect each group as an $N^2$ clique,
we emit one \emph{hub} node per group and a star of $N$ edges
(repository\,$\leftrightarrow$\,hub), turning an $O(N^2)$ blow-up into $O(N)$;
parallel Louvain (PLM) then clusters the resulting bipartite repository--hub
graph, and each community's canonical representative (hubs dropped) becomes the
deforked project $P$. The hub device is what keeps the encoding linear: an
explicit shared-evidence node replaces the quadratic clique of pairwise links.

\subsection{The project mega-cluster (Exp.~D1)}
\label{sec:deforking-cap}
The hub--star graph still over-merges. Uncapped, the largest community absorbs
$443{,}010$ repositories and $9$ communities exceed $10^5$ members
(Table~\ref{tab:deforkcap}, top row). The cause mirrors the identity case:
a small number of \emph{trivially shared} commit groups (empty or
initial commits, identical \texttt{LICENSE}/CI scaffolding committed verbatim
across unrelated repositories) act as giant hubs that bridge otherwise-disjoint
fork families. The group-size distribution is steep: $2.74$M groups
($5.3\%$) exceed $1{,}000$ repositories and the largest single group names
$267{,}200$.

\paragraph{A size cap is the cheap first layer.}
Dropping groups larger than a cap \textsc{C} removes the boilerplate bridges
while genuine fork families stay connected through their many smaller shared
groups: the precision-improving, recall-safe move that parallels neutralizing
generic identity attributes before unioning. We swept \textsc{C}; the cap is
applied inside the graph builder (skipping a group once its field count exceeds
\textsc{C}, before any node id is assigned), which is essentially free and
avoids a separate field-splitting pre-filter that was pathological on the
$10^5$-field giant lines. Table~\ref{tab:deforkcap} reports the outcome.

\begin{table}[t]
\centering\small
\caption{Deforking size-cap sweep (V2604). ``communities'' counts canonical
representatives; ``families'' counts communities with $\ge2$ repositories (a
proxy for genuine fork families); ``max'' is the largest community;
``in clusters $\ge10$'' counts repositories in communities of $\ge10$ members;
the remaining rows count communities in each size bin.}
\label{tab:deforkcap}
\begin{tabular}{@{}lrrr@{}}
\toprule
metric & uncapped & \textbf{C${=}250$} & C${=}500$ \\
\midrule
communities          & $190{,}273{,}566$ & $205{,}350{,}516$ & $201{,}029{,}443$ \\
families ($\ge2$)     & $14{,}017{,}694$  & $\mathbf{14{,}350{,}858}$ & $14{,}269{,}737$ \\
max cluster          & $\mathbf{443{,}010}$ & $\mathbf{183{,}654}$ & $183{,}957$ \\
$\ge100$             & $74{,}483$ & $77{,}624$ & $77{,}921$ \\
$\ge1{,}000$          & $5{,}623$  & $\mathbf{2{,}230}$ & $3{,}635$ \\
$\ge10{,}000$         & $334$       & $\mathbf{65}$ & $85$ \\
$\ge100{,}000$        & $9$        & $\mathbf{1}$ & $2$ \\
in clusters $\ge10$   & $58{,}289{,}235$ & $\mathbf{42{,}744{,}962}$ & $47{,}187{,}115$ \\
groups dropped       & ---         & $5.92$M ($11.4\%$) & $4.13$M ($8.0\%$) \\
\bottomrule
\end{tabular}
\end{table}

\begin{figure}[t]
\centering
\includegraphics[width=0.82\linewidth]{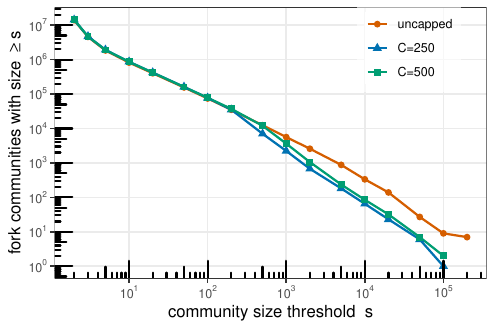}
\caption{Community-size tail before and after capping (V2604). The vertical axis
counts fork communities at or above each size threshold; capping at
\textsc{C}${=}250$ collapses the large-size tails while leaving the small-family
bulk untouched.}
\label{fig:capsweep}
\end{figure}

The cap is decisive on every mega-cluster metric and costs no families
(Figure~\ref{fig:capsweep}).
At \textsc{C}${=}250$ the largest community shrinks $443{,}010\to183{,}654$
($-58.5\%$), the $\ge10^5$ tail collapses $9\to1$, the $\ge10^4$ tail
$334\to65$, and $15.5$M repositories leave the $\ge10$ over-merge regime
($58.3\text{M}\to42.7\text{M}$). The count of genuine multi-repository
families rises only modestly ($14.018\text{M}\to14.351\text{M}$, $+2.4\%$):
the cap severs the boilerplate bridges, splitting the few over-merged
megaclusters into their constituent families without shattering the genuine
ones. \textsc{C}${=}250$ dominates \textsc{C}${=}500$ on every
mega-cluster metric \emph{and} preserves more families ($14.351$M vs.\
$14.270$M), so \textsc{C}${=}250$ is the preferred cap. Modularity is $0.999$
at both caps. The total community count \emph{rises} under capping
($190.3\text{M}\to205.4\text{M}$ at \textsc{C}${=}250$) because
dropping the boilerplate groups disconnects repositories whose \emph{only} link
into a community ran through one of those dropped hubs: such a repository falls
back to a singleton community of its own, which adds canonical representatives
even as the mega-clusters collapse: more communities, but smaller and cleaner
ones.

On reading Table~\ref{tab:artifacts}: each \texttt{p2PFull} map is deduplicated
to one row per repository over the same universe, so all three carry
$268{,}855{,}224$ rows (the number of distinct repositories covered). The
uncapped \texttt{p2PFull.V2604.s} deforks these into the $190{,}273{,}566$
communities above; the capped maps carry more communities, which rise under
capping for the singleton-fallback reason just given. The community counts are
the populations to compare across the capped and uncapped maps.

\subsection{The residual is genuine shared history, not an artifact (Exp.~D2)}
\label{sec:deforking-bridge}
A cap cannot finish the job, but (unlike the identity case) neither can a
structural cut, and that is the point. Both \textsc{C}${=}250$ and
\textsc{C}${=}500$ bottom out at the \emph{same} largest community ($183{,}654$
vs.\ $183{,}957$): the residual is held together not by the giant boilerplate
groups a cap removes but by a long tail of \emph{small} ($\le\textsc{C}$)
shared-commit groups. Removing a few oversized bridges is the standard remedy when a graph over-merges
through a handful of high-degree connectors, so we expected the principled fix to
be structural: cut the small set of high-betweenness articulation nodes that hold
the residual together. That same structural cut dissolves the analogous
author-identity mega-cluster in the companion construction~\cite{aliasingmethod},
which made it the natural thing to try here. We ran exactly that diagnostic; it
tells a different and more useful story for projects than for identities.

\paragraph{Structural diagnosis.}
We reconstruct the induced bipartite subgraph of the largest \textsc{C}${=}250$
community in NetworKit: its $183{,}654$ repositories plus every $\ge2$-member hub
touching them give $670{,}127$ nodes ($486{,}473$ shared-commit hubs) and
$10.28$M star edges, forming a \emph{single} connected component with $26{,}762$
articulation points. We compute sampled betweenness (Brandes, $4{,}000$ samples).
The load-bearing nodes are overwhelmingly \emph{repositories, not hubs}: across
the top $500$ by betweenness, project nodes carry $85.4\%$ of the mass (sum
$1.06$ vs.\ $0.18$ for hubs) and the entire top $30$ are repositories:
\texttt{qtwebengine-chromium} ($0.110$), Debian's \texttt{firefox} and
\texttt{thunderbird} mono-repos, Android \texttt{platform/external/qemu}, the
\texttt{repo.or.cz} \texttt{gcc} mirror triplet. These are aggregator and
vendoring repositories that genuinely absorbed many upstream histories, not
boilerplate scaffolding the cap missed.

\paragraph{The residual does not yield to a cut.}
We removed the top-$k$ betweenness nodes (any kind) and recomputed connected
components (Table~\ref{tab:deforkbridge}). The residual is strikingly
robust: cutting the top $1{,}000$ nodes ($724$ aggregator repositories and
$276$ hubs) shrinks the giant only $183{,}654\to145{,}661$ ($-20.7\%$) and
spawns just $531$ new families. This is the opposite of the author mega, which a
betweenness gate shattered into sub-$1$k fragments. The reason is that the
project residual is \emph{real}: Chromium is vendored into Qt, Qt and Mozilla
into Debian, half of upstream into the Android platform tree, and these repos
share genuine commits with each member they link. There is no small set of
bridges to sever: the over-merge is not an artifact but a densely
woven web of true shared history, and the high-betweenness nodes are the
aggregator repositories one would never silently delete.

We are careful about what this cut robustness does and does not prove. Robustness
to a top-$k$ betweenness cut is \emph{necessary} for the residual to be genuine but
not by itself \emph{sufficient}: a diffuse over-merge held together by many small
shared boilerplate groups (rather than a few severable bridges) would also resist
a sparse node cut, so cut robustness alone cannot distinguish ``genuine vendored
history'' from ``diffuse boilerplate over-merge.'' What discriminates the two here
is the \emph{identity} of the load-bearing nodes, rather than the fact that they
resist removal: the top-betweenness nodes are concretely identifiable aggregator and
vendoring repositories (\texttt{qtwebengine-chromium}, the Debian \texttt{firefox}/%
\texttt{thunderbird} mono-repos, the Android \texttt{platform/external} tree, the
\texttt{repo.or.cz} \texttt{gcc} mirrors) that demonstrably share \emph{real}
commits with the members they link, verifiable case by case against the upstream
projects. The cut robustness rules out the few-bridges artifact; the manual
identification of these nodes as true vendoring relationships is what establishes
the residual is genuine. The two together, not the cut alone, support the claim.

\begin{table}[t]
\centering\small
\caption{Structural-bridge diagnostic on the residual \textsc{C}${=}250$
mega-cluster ($183{,}654$ repositories, $670{,}127$-node bipartite subgraph,
one connected component). We remove the top-$k$ betweenness nodes and recompute
connected components; ``max'' is the largest surviving component (repositories),
and ``cut p/h'' splits the $k$ removed nodes into projects/hubs.}
\label{tab:deforkbridge}
\begin{tabular}{@{}lrrrr@{}}
\toprule
cut top-$k$ & components & max & families ($\ge2$) & cut p/h \\
\midrule
$0$ (residual) & $1$     & $183{,}654$ & $1$   & $0/0$ \\
$20$           & $24$    & $178{,}284$ & $13$  & $20/0$ \\
$100$          & $114$   & $168{,}890$ & $55$  & $91/9$ \\
$500$          & $881$   & $157{,}014$ & $252$ & $384/116$ \\
$1{,}000$      & $2{,}960$ & $\mathbf{145{,}661}$ & $531$ & $724/276$ \\
\bottomrule
\end{tabular}
\end{table}

\begin{figure}[t]
\centering
\includegraphics[width=0.82\linewidth]{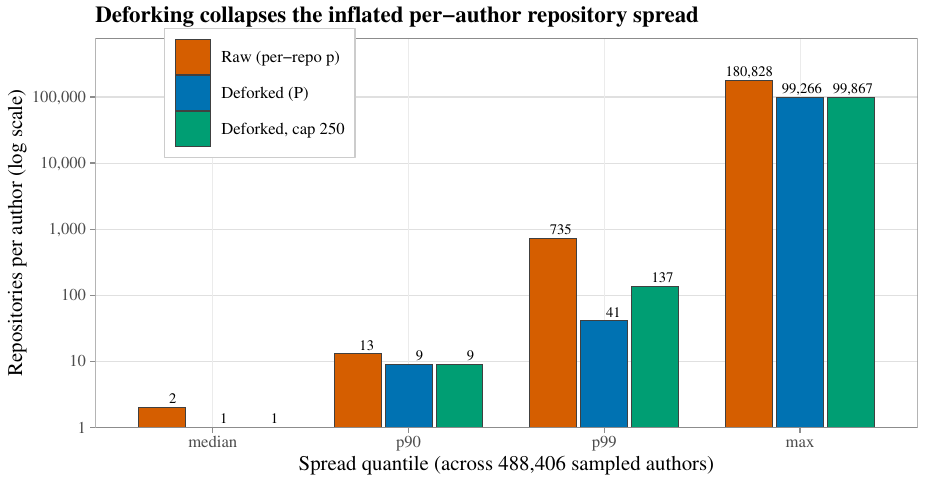}
\caption{Per-author repository spread by quantile (1/128 hash sample, $488{,}406$
authors; log scale). The raw per-repo count $p$ is fork-inflated ($p99{=}735$),
deforking collapses the tail ($p99{=}41$), and the \textsc{C}${=}250$ cap restores
a measured amount ($p99{=}137$) without reopening the mega-cluster.}
\label{fig:spread}
\end{figure}

\paragraph{Effect on the developer spread signal (Exp.~D3).}
The empirical case for a deforking choice is not the cluster-size table but how
much it moves the downstream identity signals it exists to protect. The core such
signal is \emph{spread}: the number of distinct projects a resolved author
touches, which separates real multi-project developers from artifacts and is a
primary feature of the alias graph~\cite{aliasingmethod}. We measured
it directly by joining the raw commit$\to$repository map with the resolved-author
map over a $1/128$ hash sample of authors ($488{,}406$ authors,
$43.75$M author--repository pairs), then counting distinct projects per author
under three project axes: raw repository $p$ (no deforking), the currently
distributed uncapped $P$, and the capped $P$ (\textsc{C}${=}250$). Table~\ref{tab:deforkspread}
reports the result, and it is decisive. Uncapped deforking removes
\emph{$95.1\%$} of all author spread ($43.75$M$\to2.15$M pairs); mean spread
collapses $89.6\to4.4$ and the count of prolific authors with $>\!1{,}000$
distinct projects is annihilated, $4{,}156\to15$. The collapse is concentrated in
the tail: the median holds at $1$--$2$ and p90 at $9$--$13$ across all three axes,
so it is the prolific multi-project authors (p99 $735\to41$, the
$>\!1{,}000$-project count $4{,}156\to15$) whose spread the megacluster destroys,
not the typical developer's (Figure~\ref{fig:spread}). This is the boilerplate-hub
over-merge of Exp.~D1 seen from the developer side: authors are fused across
unrelated projects bridged by trivial shared commits, erasing the
multi-project signal the identity model depends on. The \textsc{C}${=}250$ cap
restores most of it: total spread $2.15$M$\to18.7$M ($+769\%$ over uncapped),
mean $38.2$, prolific-author count back to $2{,}213$, while still collapsing
genuine forks (its mean $38.2$ sits between the fork-inflated raw $89.6$ and the
over-collapsed uncapped $4.4$, as it should). The correction is almost
purely \emph{restorative}: relative to uncapped, the cap raises spread for
$9{,}671$ authors ($1.98\%$ of the sample) and lowers it for only $341$
($0.07\%$), a $28{:}1$ asymmetry confirming the cap gives developers back
distinct projects the megacluster wrongly merged rather than fragmenting real
fork families.

\begin{table}[t]
\centering\small
\caption{Developer spread (distinct projects per resolved author) under three
project axes, over a $1/128$ author hash sample ($488{,}406$ authors). Raw
repository $p$ is fork-inflated, uncapped $P$ over-collapses, and the
\textsc{C}${=}250$ cap restores most of the signal while still collapsing genuine
forks.}
\label{tab:deforkspread}
\begin{tabular}{@{}lrrr@{}}
\toprule
per-author spread & raw $p$ & uncapped $P$ & cap250 $P$ \\
\midrule
mean                 & $89.6$       & $4.4$       & $38.2$ \\
median               & $2$          & $1$         & $1$ \\
p90                  & $13$         & $9$         & $9$ \\
p99                  & $735$        & $41$        & $137$ \\
max                  & $180{,}828$  & $99{,}266$  & $99{,}867$ \\
authors $>100$       & $11{,}418$   & $885$       & $5{,}380$ \\
authors $>1{,}000$   & $4{,}156$    & $\mathbf{15}$ & $2{,}213$ \\
\midrule
total (M pairs)      & $43.75$      & $2.15$      & $18.67$ \\
\bottomrule
\end{tabular}
\end{table}

\subsection{External validation against GitHub's declared fork graph (Exp.~D4)}
\label{sec:deforking-val}
Exps.~D1--D3 argue the cap is right \emph{internally} (cluster geometry,
structural residual, downstream spread). We now validate the deforking
\emph{externally} against a source that is independent of commit sharing:
GitHub's own declared fork relation. GitHub records, for every fork created
through its UI, a direct parent\,$\to$\,child edge; these surface in the public
event stream as \texttt{ForkEvent}s, which we extract from our GHArchive mirror
(2011--2026) into \texttt{ghaForkMap}: $135{,}268{,}473$ deduplicated named
child\,$\to$\,parent edges over $153.07$M distinct GitHub repositories, each
mapped into the WoC project namespace (lowercase, \texttt{/}$\to$\texttt{\_}).
Declared forking and commit-based deforking are methodologically disjoint (one
reads social metadata, the other reads shared SHAs), so their agreement is a
genuine cross-check rather than a tautology, in the same spirit as the
ALFAA/Bock ground truths for identities.

\paragraph{Edge-level agreement.}
For each declared fork edge we look up the deforked project $P$ of both
endpoints. Table~\ref{tab:deforkval} (left) gives the breakdown. The four
edge-cell rows are not a partition and do not sum to $100\%$: the
child-not-in-WoC and parent-not-in-WoC conditions are independent and overlap
(an edge can have both endpoints missing), so we report the cells as marginal
shares of the $135.27$M declared edges rather than as disjoint classes, while the
agreement, root, and rep figures are computed on the conditioned subpopulations
stated in each row. Two thirds of
declared forks have a child WoC never ingested ($65.19\%$): the empty or
never-pushed forks GitHub creates by the million. A further $11.04\%$ have
a parent WoC never ingested (a completeness signal, below). The decisive cell is
the conditional: \emph{restricted to the $46.10$M edges where both endpoints
exist in WoC, commit-based deforking places parent and child in the same
deforked project $99.01\%$ of the time}. The residual $0.99\%$ are declared
forks that WoC deliberately keeps separate (forks that diverged completely or
never shared a commit), which is the correct call for a commit-based method.

\emph{What this agreement does and does not validate.} The $99.01\%$ measures
\emph{recall} of the declared-fork relation: of repository pairs GitHub asserts to
be forks, how often deforking keeps them together. It confirms the construction
does not erroneously \emph{split} true forks. It does \emph{not} validate
over-merge \emph{precision} (whether deforking welds together repositories that
are \emph{not} forks of one another) because the declared graph offers no negative
fork edges to test against (GitHub records forks, not non-forks). Over-merge is the
characteristic failure mode of shared-commit clustering and the very thing the
size cap and the structural diagnostic (Exps.~D1--D2) target, so it is addressed
\emph{internally}, by the cap sweep, the residual-genuineness inspection, and the
spread-restoration measurement (Exp.~D3), rather than by this external check. The
two are complementary: D4 bounds under-merge against an independent source; D1--D3
bound and characterize over-merge. Neither alone certifies the map.

\paragraph{Root-level agreement.}
We also resolved each fork chain to its ultimate GitHub root by pointer-stepping
\texttt{ghaForkMap} to a fixed point ($134.58$M children) and asked whether the
root and the child land in the same deforked project. Of the $45.87$M children
whose root is also in WoC, $98.91\%$ share the child's deforked group, and in
$56.79\%$ ($26.05$M) the WoC canonical representative \emph{is literally the
ultimate GitHub root repository name}: deforking rediscovers GitHub's own root
without ever consulting the fork button. Only $1.09\%$ ($499{,}663$ children)
land in a different group; we inspected all of them. They split cleanly:
$30.3\%$ are both-singleton pairs (GitHub declares a fork but the two repos share
no commit in WoC: empty or scratch-rewritten forks, student-assignment and
\texttt{fork-demo} tutorial repositories), and the bulk of the remainder are
cases where \emph{WoC is more correct than GitHub's recorded chain root}: e.g.\
a fork of Swift whose WoC group is \texttt{apple\_swift} while the GHArchive
chain root is only an intermediate fork whose own \texttt{ForkEvent} predates the
public window. The single largest divergent pair ($83{,}461$ children) is a
collision of two residual boilerplate megaclusters made entirely of
\texttt{fork-demo} tutorial repositories; its root,
\texttt{gitlab.com\_natsu1978\_caf\_msm\_3.18}, is already on our
auto-curated bad-project exclusion list (Section~\ref{sec:deforking-cap}'s
boilerplate, seen from the metadata side). Net: where a real upstream is
involved, declared and commit-based forking agree or commit-based is the more
accurate of the two.

\begin{table}[t]
\centering\small
\caption{External validation of deforking against GitHub's declared fork graph
(\texttt{ghaForkMap}, $135.27$M edges). Left: edge-level breakdown, whose
load-bearing number is the \emph{conditional} agreement of $99.01\%$ once both
endpoints are known to WoC. Right: completeness byproduct, the declared fork
roots WoC never ingested, with a live-probe sample.}
\label{tab:deforkval}
\begin{tabular}{@{}lr@{}}
\toprule
edge cell (of $135.27$M) & share \\
\midrule
child not in WoC (empty forks)      & $65.19\%$ \\
parent not in WoC (completeness)    & $11.04\%$ \\
both in WoC, \emph{same} group      & $33.74\%$ \\
both in WoC, separate (true diverge)& $0.34\%$ \\
\midrule
\textbf{agreement} $\mid$ both in WoC & $\mathbf{99.01\%}$ \\
root${=}$rep $\mid$ child in WoC      & $98.91\%$ \\
rep \emph{is} GitHub root name        & $56.79\%$ \\
\bottomrule
\end{tabular}\hfill
\begin{tabular}{@{}lr@{}}
\toprule
never-ingested fork roots & count \\
\midrule
distinct declared roots   & $18.48$M \\
\quad never ingested by WoC & $4.47$M ($24.2\%$) \\
\quad on bad-project list   & $4{,}909$ ($0.11\%$) \\
\midrule
live-probe sample          & $1{,}500$ \\
\quad alive (\texttt{ls-remote}) & $943$ \\
\quad alive, HEAD not in WoC & $631$ \\
\bottomrule
\end{tabular}
\end{table}

\paragraph{Completeness byproduct.}
The $11.04\%$ parent-not-in-WoC cell is not deforking error but a
\emph{completeness} measurement, the projects GitHub knows are fork roots but
WoC never ingested. There are $4{,}469{,}886$ such roots ($24.2\%$ of $18.48$M
distinct declared roots). To separate genuine gaps from deliberate exclusions we
cross-checked them against the pipeline's bad-project lists (gitbombs,
spoofed-author repos, and the $39.5$M-entry declared-fork exclusion
set~\cite{gousios2013ghtorrent}): only
$4{,}909$ ($0.11\%$) appear, and those are low-fork artifacts. The rest are real
upstreams that were never cloned. A $1{,}500$-root sample (top by fork
count plus a tail spread), probed with \texttt{git ls-remote} and then checked
against the WoC object database with \texttt{hasObj}, finds $943$ still live and
$631$ of those carrying a HEAD commit genuinely absent from WoC, including
high-visibility repositories such as \texttt{ollama/ollama},
\texttt{MystenLabs/sui}, and \texttt{comfyanonymous/ComfyUI}. This list feeds the
ingestion-completeness work (it is the same signal the data-track paper reports)
and is, to our knowledge, the first fork-graph-anchored estimate of WoC's
upstream-root coverage gap.

\subsection{Detached forks, cross-forge families, and a refreshed exclusion
list (Exp.~D5)}
\label{sec:deforking-d5}
Exp.~D4 measured \emph{agreement}; the same join exposes three further questions
that complete the deforking story: which declared forks have genuinely
\emph{detached} from their parent, how often fork families span more than one
forge, and whether the WoC fork-exclusion list (used to skip redundant
re-collection) is current. All figures in this subsection are computed over the
\emph{uncapped} map \texttt{p2PFull.V2604.s} (the version WoC currently
distributes) rather than the recommended \textsc{C}${=}250$ map, because the
exclusion list and cross-forge inventory must match the map currently in
circulation. A consequence is that the group-size bins below (and the cross-forge
and root-forge counts) inherit the uncapped map's over-merge: the large-group tail
in particular conflates genuine families with the boilerplate clusters the cap
would trim, which we flag where it matters and would recompute if the
capped map were adopted.

\paragraph{Detached forks.}
Restricting to the $46{,}098{,}601$ declared-fork edges whose \emph{both}
endpoints are in WoC, $455{,}550$ ($0.99\%$) are kept in \emph{different}
canonical groups by commit-based deforking ($P_{\text{child}}\neq
P_{\text{parent}}$; abbreviated $P_c,P_p$ in the released
\texttt{detached\_forks} schema). In other words $99\%$ of in-WoC GitHub forks defork
together with their parent, and the $\sim1\%$ that do not are forks whose commit
history no longer overlaps the parent in WoC, the operational definition of a
``hard fork.'' Table~\ref{tab:deforkd5} (left) bins these by the size of the
child's own deforked group. Nearly half ($46.1\%$) are \emph{lone divergences}
($|P_{\text{child}}|{=}1$): the fork shares no ingested commit with the parent,
which is either a true history rewrite/squash or a completeness gap in the shared
base (the two are separated by the GHArchive force-push analysis, future work).
The cleanest positive signal (the child is itself a deforked representative of a
non-mega family, $P_{\text{child}}{=}\text{child}$ with $|P_{\text{child}}|\le
1000$) holds for $240{,}441$ forks: these are independent origins that grew
their own downstream fork trees. The large-group tail ($21.6\%$ in groups
$>$$10$k) is mixed: some are legitimately massive families
(\texttt{airbnb/javascript}), and the top entries sit in the chromium-in-Qt
aggregator whose genuine vendored core is the \textsc{C}${=}250$ residual analyzed
in Exp.~D2 (\texttt{qtwebengine-chromium}, a real vendoring relationship, not a
boilerplate over-merge), wrapped, under the \emph{uncapped} map used for
this join, in the cap-removable periphery the size cap of Exp.~D1 trims. The bin is
therefore structurally heterogeneous (true independent origins, genuine vendored
aggregators, and cap-removable boilerplate together), so we read it qualitatively
rather than as a count of detachments.

\paragraph{Cross-forge families and roots off GitHub.}
Because commit-based deforking is forge-agnostic (it clusters by shared commits,
not by any one forge's fork pointer), it sees fork relationships GitHub's graph
cannot. Over $190{,}273{,}566$ deforked groups, $14{,}017{,}694$ ($7.37\%$) are
fork families (size${>}1$). Of those, $775{,}420$ ($5.53\%$) are
\emph{cross-forge}: members on more than one forge, i.e.\ the same codebase
mirrored across GitHub, GitLab, Bitbucket, SourceForge, and others, a population
invisible to any single forge's fork metadata. And $215{,}672$ ($1.54\%$ of
families) have a canonical root that is \emph{not on GitHub at all}
(Table~\ref{tab:deforkd5}, right): GitLab ($0.61\%$), Bitbucket ($0.48\%$),
SourceForge ($0.15\%$), then Gitorious, HuggingFace, salsa.debian, and Launchpad,
with a long tail of $\sim\!30$ further forges (Drupal, Codeberg, framagit,
\ldots) aggregated as ``other'' in Table~\ref{tab:deforkd5}. So while
$\sim98.5\%$ of forked projects do trace to a GitHub
origin, one in $65$ fork families is rooted on another forge and would be
misattributed by a GitHub-only fork graph.

\paragraph{Refreshed exclusion list.}
The pipeline skips re-collecting declared forks via a static list
(\texttt{woc.pm}'s \texttt{addForks()}), sourced from a GHTorrent-era
dump~\cite{gousios2013ghtorrent} of
$39{,}475{,}737$ repositories. The GHArchive fork map supplies a current
replacement: $134{,}575{,}790$ distinct declared-fork children (through
2026-06, $3.4\times$ larger). We subtract the detached forks (those should keep
being updated because they carry independent development), yielding
$134{,}132{,}789$ repositories to exclude (\texttt{excludeForksGHA.gz}), or
$134{,}335{,}349$ under a conservative variant that keeps only the $240{,}441$
detached children that became their own multi-member family. The subtraction is
over \emph{distinct children}, not edges: the $455{,}550$ detached
\emph{edges} of Table~\ref{tab:deforkd5} collapse to $443{,}001$ distinct
detached child repositories (a child can be a declared fork of more than one
parent, so one detached child can contribute several edges), and it is those
$443{,}001$ children (of which $240{,}441$ form their own independent
family) that are removed from the exclusion universe.

\begin{table}[t]
\centering
\small
\caption{Exp.~D5. Left: detached forks ($P_{\text{child}}\neq P_{\text{parent}}$,
both in WoC; $455{,}550$ edges) by child-group size. Right: canonical-root forge
of the $14{,}017{,}694$ fork families; the column is exhaustive and sums to the
total. $^{\dagger}$``other'' aggregates the $25{,}199$-family tail below
\texttt{git.launchpad.net}.}
\label{tab:deforkd5}
\begin{tabular}{lrr}
\toprule
child-group size $|P_{\text{child}}|$ & edges & \% \\
\midrule
$1$ (lone divergence)        & $210{,}030$ & $46.1$ \\
$2$--$10$ (small family)     & $74{,}136$  & $16.3$ \\
$11$--$100$                  & $37{,}769$  & $8.3$ \\
$101$--$1000$                & $21{,}547$  & $4.7$ \\
$1001$--$10{,}000$           & $13{,}626$  & $3.0$ \\
$>$$10{,}000$ (mega/over-merge) & $98{,}442$ & $21.6$ \\
\bottomrule
\end{tabular}
\hfill
\begin{tabular}{lrr}
\toprule
root forge (families) & count & \% \\
\midrule
github            & $13{,}802{,}022$ & $98.46$ \\
gitlab.com        & $85{,}789$ & $0.61$ \\
bitbucket.org     & $66{,}983$ & $0.48$ \\
sourceforge.net   & $21{,}143$ & $0.15$ \\
gitorious.org     & $4{,}884$  & $0.03$ \\
huggingface.com   & $4{,}809$  & $0.03$ \\
salsa.debian.org  & $3{,}680$  & $0.03$ \\
git.launchpad.net & $3{,}185$  & $0.02$ \\
other$^{\dagger}$ & $25{,}199$ & $0.18$ \\
\bottomrule
\end{tabular}
\end{table}

\subsection{Threats to validity}
\label{sec:deforking-threats}
Seven limitations bound the construction and validation claims. \textbf{(a) GHArchive
window.} The declared fork graph is reconstructed from the public event stream,
which begins in 2011. Forks created before 2011, and forks of repositories whose
defining \texttt{ForkEvent} predates the window, are invisible to
\texttt{ghaForkMap}; this biases the ``parent not in WoC'' cell and the
root-resolution step (a chain can appear rooted at an intermediate fork whose
own \texttt{ForkEvent} we never saw), so both the completeness estimate and the
$56.79\%$ ``rep is GitHub root'' figure are conditioned on the observable window
rather than on all-time GitHub history. \textbf{(b) Namespace mapping.} GitHub
repository names are mapped into the WoC project namespace by lowercasing and
replacing \texttt{/} with \texttt{\_}; this can collide distinct repositories
(e.g.\ owners differing only in case, or names containing literal underscores)
and so can spuriously raise or lower edge agreement at the margin. We have not
quantified the collision rate; it is a known source of noise in the $99.01\%$.
\textbf{(c) Conditional agreement.} The headline $99.01\%$ is computed only on
the $46.10$M edges whose \emph{both} endpoints are present in WoC, roughly
$34\%$ of the $135.27$M declared edges. It is therefore an agreement
\emph{conditional on ingestion}, not an unconditional accuracy over all declared
forks; the two-thirds of edges with a missing endpoint are characterized
separately (empty forks, completeness gaps) but cannot be checked for agreement.
\textbf{(d) Modularity.} The reported modularity of $\sim\!0.999$ should not be
read as a pure quality signal: the bipartite repository--hub star encoding makes
the graph extremely sparse and locally tree-like, which inflates modularity
mechanically for almost any reasonable partition. We use it only as a sanity
check that the partition is non-degenerate, not as evidence of cluster
correctness; the substantive evidence is the external agreement and the
spread-restoration measurement. \textbf{(e) Clustering stochasticity.} Parallel
Louvain is order- and seed-dependent, so the exact membership of borderline
clusters near the cap threshold is not guaranteed reproducible to the individual
repository; we fix the seed and report the cap as the tunable knob, and the
aggregate external-agreement and spread-restoration results are stable across
runs, but per-repository boundary assignments in the mega-cluster tail are not.
\textbf{(f) Spread-restoration sample.} The spread effect is measured over a
$1/128$ author hash sample ($488{,}406$ authors); aggregate statistics are
representative, but the extreme-spread tail is sampled sparsely, and
$\textsc{C}{=}250$ is one operating point on the cap--recall tradeoff rather than
a tuned optimum. \textbf{(g) Forge attribution.} The root forge of a family is
inferred from the WoC project-name prefix (\texttt{gitlab.com\_},
\texttt{bitbucket.org\_}, \ldots); repositories whose names carry no host prefix
default to GitHub, so the cross-forge ($5.53\%$) and root-off-GitHub ($1.54\%$)
counts are lower bounds wherever a non-GitHub forge omitted its prefix.

\paragraph{Disposition.}
The two experiments draw a clean line. The size cap (Exp.~D1) removes the
boilerplate-hub over-merge decisively and recall-safely, and is the change we
recommend (\textsc{C}${=}250$). The structural diagnostic (Exp.~D2) shows
the surviving residual is \emph{not} a defect of the same kind: it is true
vendored history that no bridge cut dissolves, so deleting it would be a
precision loss, not a gain. We therefore recommend that the uncapped deforked map
\texttt{p2PFull.V2604.s} WoC currently distributes be left intact; the capped maps
\texttt{p2PFull.V2604.cap\{250,500\}.s} and the full betweenness/articulation
analysis are released in the replication package as the evidence base for the
\textsc{C}${=}250$ cap and for treating the aggregator residual as genuine rather
than surgically cutting it.

\paragraph{Future work.}
Four directions remain open. (i)~\emph{Adoption.} The \textsc{C}${=}250$ map is
the deforking we recommend; adopting it in WoC would require repointing the
collection's output through the cross-version rank-stabilization step so that
canonical representatives stay comparable to prior releases. We will offer this
change to the World of Code maintainers; until such a migration is validated, the
uncapped map remains the one in circulation. (ii)~\emph{Downstream re-measurement.}
Exp.~D3 settles the spread half of this: the cap restores $769\%$ of the
author-spread signal that uncapped deforking destroys, which is the empirical
case for adoption. What remains is to recompute the \emph{overlap}-based identity
signals~\cite{aliasingmethod} and the alias graph itself over the
capped $P$, and to confirm the spread restoration carries through to the final
per-edge classifier rather than only the raw feature. (iii)~\emph{The aggregator
residual.} Exp.~D2 shows the residual is genuine vendored history that no
structural cut separates without precision loss; splitting it, if ever desired
(e.g.\ to attribute Chromium-in-Qt-in-Debian distinctly), calls for a
\emph{semantic} policy that recognizes vendoring/mirroring relationships rather
than a topology-only cut. We regard this as low priority because the
residual is not an artifact.
(iv)~\emph{Blob-based provenance for commit-disjoint upstreams.} Shared-commit
deforking cannot link repositories that share no commits yet are genuine
upstream/downstream pairs. The clearest case is package mirrors that re-commit
released versions instead of carrying the upstream history: a CRAN-style mirror
that checks each release in as a fresh tree reproduces the same source content
under entirely new commit hashes, so the commit-sharing signal is destroyed even
though the code provenance is real. Validating, and extending, the deforking map
against \emph{shared blob} content (which survives re-committing) would both
cross-check the commit-based clusters and recover these commit-disjoint source
relationships. This is a substantial undertaking, in scale (the blob-sharing
relation is far denser than commit sharing) and in policy (blob overlap conflates
vendoring, copy-paste, and boilerplate, so it needs the same precision discipline
applied here to commits), and we leave it to future work.

\section{Availability}
\label{sec:availability}
All artifacts are released as a single self-contained bundle; no World of Code
account is required to obtain or use them (artifact schemas, keys, and row
counts are summarized in Table~\ref{tab:artifacts}): the deforking map
\texttt{p2PFull.V2604.s} and its capped variants
\texttt{p2PFull.V2604.cap250.s} / \texttt{p2PFull.V2604.cap500.s}, the refreshed
fork-exclusion list (\texttt{excludeForksGHA.gz} and its conservative variant),
the detached-fork inventory, the GHArchive fork map (\texttt{ghaForkMap.gz}),
and the GitHub-fork-graph agreement tables, each \texttt{;}-separated,
gzip-compressed, and \texttt{LC\_ALL=C} sorted, together with a replication
package that regenerates them from the WoC shared-commit relation and the
GHArchive \texttt{ForkEvent} mirror. Because the maps are multi-gigabyte, the
data artifacts are hosted as a Hugging Face dataset (which scales past the
per-file limits of a code host and mints a citable DataCite DOI), with the
replication code mirrored on GitHub and cross-linked from the dataset card. We
will also offer the identical bundle to the World of Code maintainers, so that
existing WoC users could obtain it through the channels they already use, should
the maintainers choose to adopt it. The data is released under \textsc{CC-BY-4.0} and the
replication code under the \textsc{MIT} license. \emph{(Hugging Face dataset:
\texttt{TODO/woc-deforking-v2604}; DOI to be minted on camera-ready.)}

\begin{table*}[t]
\centering\small
\caption{Released artifacts (WoC \texttt{V2604}). All files are
\texttt{;}-separated, gzip-compressed, and \texttt{LC\_ALL=C} sorted on the
first column. The \texttt{rows} column reports each file's line count: one row
per declared edge for the GHArchive fork map, and one row per repository for the
exclusion lists and the deduplicated \texttt{p2PFull} maps.}
\label{tab:artifacts}
\begin{tabular}{@{}llllr@{}}
\toprule
artifact & schema (columns) & key & sharding & rows \\
\midrule
\texttt{p2PFull.V2604.s}        & \texttt{p;P} (repository; deforked project) & \texttt{p} & 32 author-hash shards / single merged file & $268{,}855{,}224$ \\
\texttt{p2PFull.V2604.cap250.s} & \texttt{p;P} (cap \textsc{C}${=}250$)       & \texttt{p} & single file                        & $268{,}855{,}224$ \\
\texttt{p2PFull.V2604.cap500.s} & \texttt{p;P} (cap \textsc{C}${=}500$)       & \texttt{p} & single file                        & $268{,}855{,}224$ \\
\texttt{ghaForkMap.gz}          & \texttt{child;parent} (declared fork edge)  & \texttt{child} & single file                    & $135{,}268{,}473$ \\
\texttt{excludeForksGHA.gz}     & \texttt{child} (repository to skip)         & \texttt{child} & single file                    & $134{,}132{,}789$ \\
\texttt{excludeForksGHA.consv.gz} & \texttt{child} (conservative variant)     & \texttt{child} & single file                    & $134{,}335{,}349$ \\
\texttt{detached\_forks.gz}     & \texttt{child;parent;$P_c$;$P_p$;$|P_c|$;$|P_p|$} & \texttt{child} & single file              & $455{,}550$ \\
\bottomrule
\end{tabular}
\end{table*}

\section{Conclusion}
\label{sec:conclusion}
Deforking is the project-graph analogue of author-identity disambiguation: the
naive shared-history union over-merges for the same structural reason, and yields
to the same cap-then-cut fix. The released map deforks the entire WoC collection,
validates at $99.01\%$ edge agreement against GitHub's declared fork graph
(conditional on both fork endpoints being present in WoC), and, because it is
built from observed commits rather than platform metadata, recovers
cross-forge and detached fork families that no single platform graph can see,
making the map a more complete provenance substrate than any forge's own fork
records.

\bibliographystyle{ACM-Reference-Format}
\bibliography{refs}

\end{document}